\documentclass[12pt,english]{article}
\usepackage[T1]{fontenc}
\usepackage[latin1]{inputenc}
\usepackage{amsmath}
\usepackage{amssymb}
\usepackage{mathrsfs}
\usepackage{amsmath,epsfig}
\usepackage{graphicx}
\makeatletter

\catcode`@=12
\usepackage{babel}
\makeatother
\newcommand {\be}{\begin{equation}}
\newcommand {\ee}{\end{equation}}
\newcommand {\ba}{\begin{eqnarray}}
\newcommand {\ea}{\end{eqnarray}}
\begin{document}
\thispagestyle{empty}
\begin{flushright}
IPM/P-2008/003\\
\today
\end{flushright}

\mbox{} \vspace{0.75in}

\begin{center}\textbf{\large Probing of $Wtb$ Anomalous Couplings via the $tW$
Channel of Single
Top Production}\\
 \vspace{0.5in} \textbf{$\textrm{Mojtaba Mohammadi Najafabadi}$\footnote{mojtaba@mail.ipm.ir} }\\
 \vspace{0.2in}
{\sl School of Particles and Accelerators, \\
IPM (Institute for Studies in Theoretical Physics and Mathematics)\\
P.O. Box 19395-5531, Tehran, Iran}\\
 \vspace{.75in}\end{center}

\baselineskip 20pt
\begin{abstract}
The potential of LHC for investigation of the $W$-$t$-$b$
vertex through the $tW$ channel of single top quark production is
studied. Unlike the other two single top quark production processes
($t-$channel and $s-$channel), the $tW$ channel provides the
possibility to study the $Wtb$ vertex without receiving
contamination from FCNC. This study has been done at parton level
but is involved the separation of signal from backgrounds when
both $W$-bosons decay to leptons. In this study $\mathcal{CP}$ is
assumed to be conserved. The $68\%$ C.L. bounds on the
non-Standard Model couplings are estimated.

\end{abstract}
\newpage

\section{Introduction}
One of the main goals of the upcoming operational LHC is to search
for new physics beyond the Standard Model (SM). Because of the
large mass of the top quark among all observed particles within
the SM, it may give a special role in the generation of masses.
Therefore, it is crucial that its interactions with other
particles be studied carefully. The deviations of the top quark
interactions from the SM predictions may represent a good way to
learn more about the nature of the electroweak symmetry breaking
\cite{beneke},\cite{tait1}.

One approach to describe possible new physics
effects is to use a model independent technique based on the
effective low energy Lagrangian . In this approach, the SM Lagrangian is modified by
adding new $SU(3)_{c}\otimes SU(2)_{L}\otimes U(1)_{Y}$ gauge
invariant operators \cite{zhang1},\cite{zhang2}:
\begin{eqnarray}
\mathcal{L}_{eff} = \mathcal{L}_{SM} +
\frac{1}{\Lambda^{2}}\sum_{i}C_{i}\mathcal{O}_{i}.
\end{eqnarray}
where $\mathcal{L}_{SM}$ is the SM Lagrangian, $\Lambda$ is the
new physics scale, $\mathcal{O}_{i}$ are dimension six operators
which are gauge invariant before electroweak gauge symmetry
breaking. $C_{i}$ are constants which represent the coupling
strengths of $\mathcal{O}_{i}$. The dimension five operators
violate the lepton number conservation.

Upon electroweak symmetry breaking and giving our attention to the
top quark, the modified $Wtb$ couplings can be expressed as \cite{zhang2}:
\begin{eqnarray}
\Gamma^{\mu}_{Wtb} = -\frac{g}{\sqrt{2}}[
\gamma^{\mu}(F_{L1}P_{L}+F_{R1}P_{R})
-\frac{i\sigma^{\mu\nu}q_{\nu}}{m_{W}}(F_{L2}P_{L}+F_{R2}P_{R})]
+ (h.c.)
\end{eqnarray}
where $g$ is the weak coupling constant, $m_{W}$ is the W-boson
mass, $q_{\nu}$ is the W-boson four-momentum. $P_{R,L} =
\frac{1\pm \gamma_{5}}{2}$ is the right-handed (left handed)
projection operator. Assuming $\mathcal{CP}$ conservation, $F_{L1,2}$ and
$F_{R1,2}$ are real form factors. 
These anomalous couplings are related to the coefficients $C_{tW\Phi}$ and $C_{bW\Phi}$
in the general effective Lagrangian by \cite{zhang2}:
\begin{eqnarray}
F_{L2} = \frac{C_{tW\Phi}\sqrt{2}m_{W}v}{g\Lambda^{2}}~~~~,~~~~F_{R2} 
= \frac{C_{bW\Phi}\sqrt{2}m_{W}v}{g\Lambda^{2}}
\end{eqnarray}

where $\Lambda$ is the scale of new physics.
At tree level of the SM, $F_{L1}
= V_{tb} \simeq 1$ and $F_{R1} = F_{L2} = F_{R2} = 0$.

At hadron colliders, top quarks can be produced in $t\bar{t}$
pairs via strong interactions and singly via electroweak
interactions. The introduced anomalous couplings have been
studied via top pair events at the LHC and Tevatron in
\cite{onfre1},\cite{onfre2},\cite{tsuno}.

In the SM single top quark events are expected to be produced via
the $t-$channel process ($bq\rightarrow q't$), the $s-$channel
process ($q\bar{q}'\rightarrow t\bar{b}$) and the $tW$-channel
process ($gb\rightarrow tW^{-}$). These three processes have
completely different kinematics and can be observed separately \cite{beneke}.
The first evidence for single top was reported by D0
experiment at Tevatron \cite{singletopd0}. At the Tevatron, the
events of the $tW$-channel can not be observed because of the very
small cross section of this process.

A complete study of the $Wtb$ anomalous couplings using the
$t-$channel and the $s-$channel for LHC and Tevatron has been
performed in \cite{lev1} which has given the following bounds on
the anomalous couplings (assuming $10\%$ systematic uncertainty):
\begin{eqnarray}
-0.094\leq F_{L2} \leq 0.34 ~~~,~~~-0.17\leq F_{R2} \leq 0.18
\end{eqnarray}
The data coming from $b\rightarrow s\gamma$ has applied very tight
constraints on the $F_{R1}$ \cite{pdg}. Thus we do not consider
$F_{R1}$ in the present study.

The $tW$-channel has almost a large cross section at the LHC
($\sim 60 ~pb$) and does not receive any contribution from FCNC
(Flavor Changing Neutral Current), therefore it can be used to
study the vertex of $W$-$t$-$b$. The aim of this article is to
investigate the sensitivity of this channel to anomalous couplings
and estimation of the possible bounds on $F_{R2}$ and $F_{L2}$.

\section{The Top Quark Width Sensitivity to the Anomalous Couplings}

The Standard Model predicts the top quark lifetime to be around $4\times 10^{-25}$ s 
which corresponds to the top quark width of 1.5 GeV. One should note that because of the 
experimental restrictions, it is very difficult to measure this very short lifetime
or the corresponding width. However, we are able to set a lower limit on the top quark 
width from the available data from Tevatron. 
In \cite{cdfwidth} an upper limit has been set on the top quark width using a likelihood fit to the
reconstructed top mass distribution. In the analysis the lepton+jets channel of $t\bar{t}$ 
candidates, in which one of two W-bosons decays to $l\nu_{l}$ while the other decays to
$qq'$, is used to reconstruct the top quark mass. Finally, the estimated upper bound on the 
top quark width is 12.7 GeV with $95\%$ C.L. This is corresponding to the 
lower limit of $5.2\times 10^{-26}$ s for the top quark lifetime.
The top quark width will not be measured very precisely at the LHC \cite{cmstdr}.

\begin{figure}
\centering
  \includegraphics[height=8cm,width=12cm]{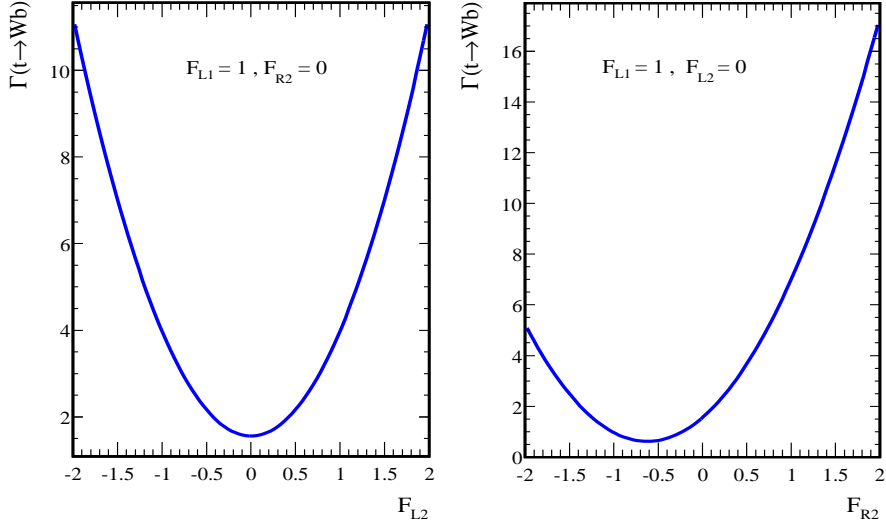}\\
  \caption{The tree level width of the top quark (in GeV) in terms of anomalous couplings}\label{width}
\end{figure}

From another side, the introduced $Wtb$ coupling, at tree level and in
the limit of $m_{b}\rightarrow 0$, leads to the following formula for the width of top quark \cite{kane},\cite{mojtaba}:

\begin{eqnarray}
\Gamma_{t\rightarrow Wb} =
\frac{G_{f}m_{t}m_{W}^{2}}{8\sqrt{2}\pi}\frac{(r^{2}-1)^{2}}{r^{4}}
[(r^{2}+2)(F^{2}_{L1}+F^{2}_{R1})\\ \nonumber
+(2r^{2}+1)(F^{2}_{L2}+F^{2}_{R2}) +6r(F_{L1}F_{R2}+F_{R1}F_{L2})]
\end{eqnarray}

where, $m_{t}$ and $m_{W}$ are top mass and W-boson mass, respectively and
$r = \frac{m_{t}}{m_{W}}$.
Fig.\ref{width} shows the top quark width as a function of
anomalous couplings $(F_{L2},F_{R2})$ at tree level.

Obviously, the
top quark width varies around 10-15 GeV when the anomalous couplings
change in a wide region $(-2.0,2.0)$. As stated above, the upper limit
on the top quark width from the reconstructed top quark invariant mass distribution
at Tevatron is around 12.7 GeV. This means that a wide region
of the anomalous couplings ($F_{L2},F_{R2}$) is allowed from the present
top quark invariant mass measurement.

In the next section the sensitivity of the cross section of the
$tW$ channel single top to the anomalous couplings is examined and
new bounds on the anomalous couplings are estimated.

\section{The $tW$ Channel Cross Section Sensitivity to the Anomalous Couplings}

\begin{figure}
\centering
  \includegraphics[height=8cm,width=12cm]{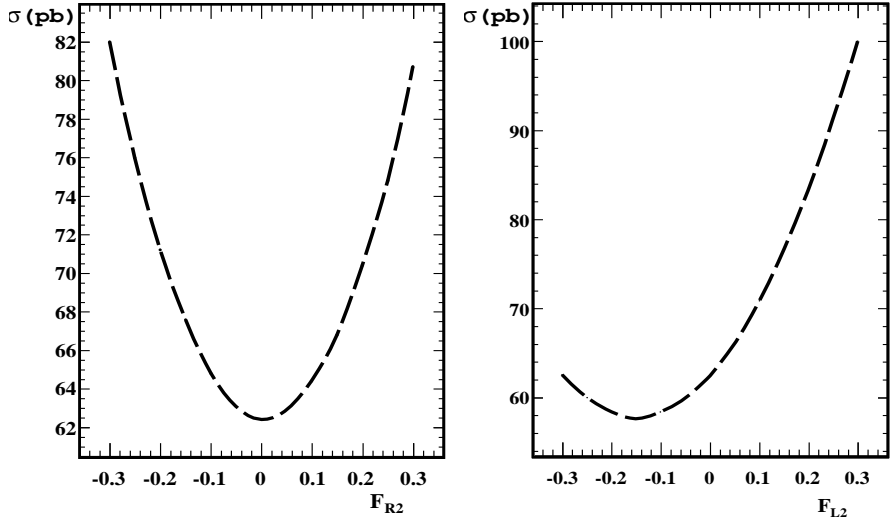}\\
  \caption{The dependency of the $tW$ channel single top production cross section on
  $F_{R2}$ when $F_{L1} = 1~,~F_{L2} = 0$ (left plot) and on $F_{L2}$ when $F_{L1} = 1~,~F_{R2} = 0$ (right plot).} \label{sigma}
\end{figure}

The dependency of the $tW$ channel of single top quark cross section on
the anomalous couplings at the LHC is presented in Fig.\ref{sigma}.
This figure has been obtained by using the CompHEP package \cite{comphep}. In 
calculation of the cross section, it is assumed that  
$m_{top}$ = 175 GeV/c$^{2}$, $m_{b}$ = 4.8 GeV/c$^{2}$ and
CTEQ6L1 is used as the proton parton distribution function.

According to CMS Collaboration full simulation results, the
relative statistical uncertainty on measurement of the cross
section $(\frac{\Delta\sigma}{\sigma})$ of the $tW$ channel
taking into account 10 fb$^{-1}$ of integrated luminosity is
9.9$\%$ \cite{cmstdr}. While ATLAS collaboration predicted 3$\%$ for this value
with 30 fb$^{-1}$ of integrated luminosity of data
\cite{atlastdr}. Therefore, the cross
section of the $tW$ channel will be measured precisely when the LHC is operational.

\begin{figure}
\centering
  \includegraphics[width=14cm,height=9cm]{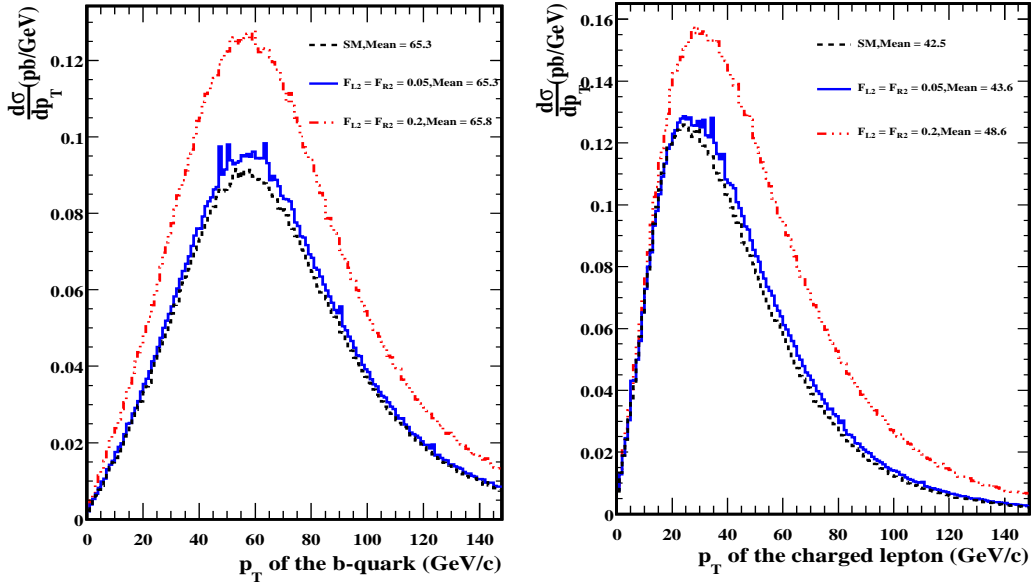}
  \caption{The transverse momentum distributions of the charged lepton and b-quark
  coming from top in the $tW$ channel single top in the SM and for different values of $F_{L2}, F_{R2}$.}\label{pt}
\end{figure}

In the $tW$ channel process the single top quark is produced via
$gb\rightarrow tW^{-}$ process. In the
di-leptonic decay mode, besides the charged lepton coming from
top quark, missing energy and b-jet, the final state contains another
charged lepton (from the real $W^{-}$-boson) with opposite sign of
the charged lepton coming from top quark.
The distribution of the transverse momentum $(p_{T} = \sqrt{p_{x}^{2} + p_{y}^{2}})$ of the charged lepton and the
b-quark from the top quark decay are shown in Fig.\ref{pt}.
According to the left plot in Fig.\ref{pt}, the transverse
momentum of the b-quark from the top quark is insensitive to the
anomalous couplings. In contrast, the transverse momentum
distribution of the charged lepton is shifted in the presence of
anomalous couplings with respect to the SM case. As it has been
shown in the right plot in Fig.\ref{pt}, the mean value of the
transverse momentum of the charged lepton is shifted toward the
large $p_{T}$ region around 6 GeV when the anomalous couplings
are $F_{L2} = F_{R2} = 0.2$. One should note that this mean value 
depends on the range of the histogram (e.g. the mean value of the $p_{T}$
distribution is increased when the range of the histogram is set to 
300 GeV/c instead of 150 GeV/c). According to the charged lepton $p_{T}$ distribution,
for very small values of $F_{L2}~,~F_{R2}$, the shift toward large $p_{T}$
region is negligible.

Modern detectors at the LHC are able to measrure the 
transverse momentum of the charged leptons (muon and electron) very precisely.
For example, the CMS detector is able to measure the $p_{T}$ of muons with 
the precision of ~$1.5\%$ (when muon is in the centeral region of the 
detector and $p_{T}\lesssim 100$ GeV/c) \cite{cmstdr1}. Therefore, a shift of around 6 GeV
might be observable which corresponds to $F_{L2} = F_{R2} = 0.2$. However,
a shift in the $p_{T}$ of the
charged lepton distribution corresponding to $F_{L2} = F_{R2} = 0.05$ is not observable .

In order to obtain a realistic estimate of the sensitivity
of the $tW$ channel single top at the LHC, one has to take into account backgrounds, detector
effects and selection cuts. Obviously, a comprehensive analysis
of all reducible backgrounds and detector effects is beyond the scope
of this study and must eventually be performed by the
experimental collaborations.
In \cite{tait2} a Monte Carlo study at parton level has been
performed which is involved the separation of signal from
backgrounds when two W-bosons decay to leptons. The most contributing
backgrounds are $t\bar{t}$ and $W^{+}W^{-}b$. The signal contains
two high $p_{T}$ charged leptons, only one jet (the b-jet coming
from top quark) and missing energy.

According to the proposed strategy for separation of signal from
backgrounds in \cite{tait2}, the charged leptons and b-jet are required to have
$p_{T} \geq 15$ GeV and to lie in the centeral region of the
detector with $|\eta| \leq 2.0$. The following angular separation
cut is applied on charged leptons and jet:
\begin{eqnarray}
 \Delta R = \sqrt{\Delta\eta^{2} + \Delta\phi^{2}} \geq 0.4.
\end{eqnarray}
where $\Delta\phi$ is separation in azimuthal angle and
$\Delta\eta$ is the difference in pseudorapidity.
After applying the above cuts, the significance at LHC
with 20 fb$^{-1}$ of integrated luminosity is
seen to be 84.9. Therefore, we use this selection strategy 
to suppress the backgrounds.

As mentioned before in this study, the CompHEP package \cite{comphep} was used to simulate the $tW$
channel single top production with anomalous couplings.
We use simple $\chi^{2}$ criterion from the transverse momentum distribution of the charged lepton
(coming from the top quark) with 20 fb$^{-1}$ of integrated luminosity to estimate the limits on anomalous $Wtb$ couplings. 
The analysis is performed after application of the mentioned cuts for backgrounds 
suppression from \cite{tait2}. The $\chi^{2}$ criterion is defined as:

\begin{eqnarray}\label{chi2}
\chi^{2}(F_{L2},F_{R2}) = \sum_{i=bins}(\frac{N^{non-SM}_{i} - N^{SM}_{i}}
{\Delta_{i}})^{2}
\end{eqnarray}

where $N^{SM}_{i}$ is the number of standard events in the i-th bin 
of the transverse momentum distribution of the charged lepton
and $N^{non-SM}_{i}$ is the number events predicted by the non-standard theory in the i-th bin
(in our case the theory with anomalous $Wtb$ couplings).

In the present study, we have taken the advantage of the fact that, for small values of 
$F_{L2}$ and $F_{R2}$, approximately the 
of the cross section of the $tW$ channel in the presence of anomalous couplings is linear in $F_{L2}$ and
$F_{R2}$ so the $N^{non-SM}_{i}$ can be written linearly in $F_{L2}$ and
$F_{R2}$ too. Therefore, $\chi^{2}$ criterion depends quadratically on anomalous couplings.

In Eq.\ref{chi2}, $\Delta_{i}$ is defined as:
\begin{eqnarray}
\Delta_{i} = N^{SM}_{i}\sqrt{\delta^{2}_{stat}+\delta^{2}_{syst}}
\end{eqnarray}
where $\delta_{stat}$ is the statistical uncertainty and $\delta_{stat}$ is the
term for including systematic uncertainties.
Systematic uncertainties from $m_{top}$, parton distribution function, 
QCD scales, luminosity measurements and etc. are important for accurate
results. However, at this stage it is difficult to give a realistic estimate of
systematics. Therefore, combined systematic uncertainties of 10$\%$ and 25$\%$ are 
taken into account.
Because of different sources of uncertainties, taking into account a combined systematic
uncertainty of 25$\%$ seems to give more realistic results.

\begin{figure}
\centering
  \includegraphics[height=8cm,width=10cm]{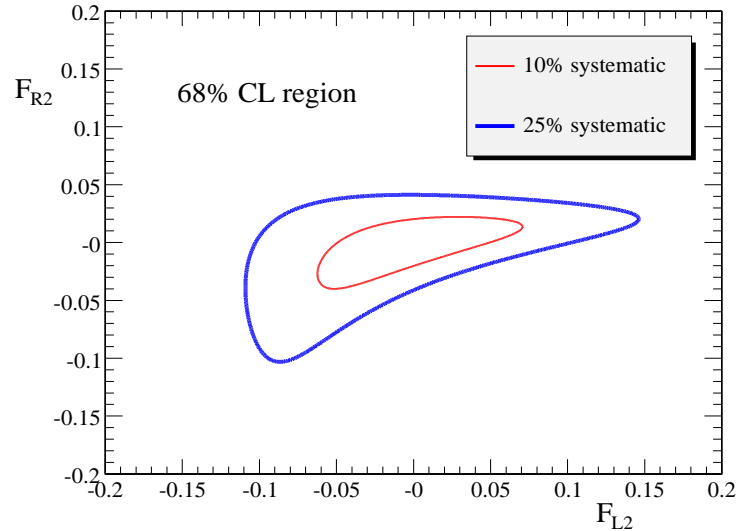}\\
  \caption{The $68\%$ confidence level regions of anomalous couplings with 10$\%$ and 25$\%$ of
  systematic uncertainties.}\label{countour}
\end{figure}

Using the $\chi^{2}$
function defined by Eq.\ref{chi2}, the 68$\%$ confidence level contours are drawn in
Fig.\ref{countour} assuming 10$\%$ and 25$\%$ of systematic uncertainties.

The $68\%$ C.L. bounds
on the non-SM couplings with different values of systematic uncertainties are given
in Table \ref{bounds}. Comparing these limits with the limits which have been estimated by using the $t-$channel
and $s-$channel of single top from \cite{lev1} (mentioned in the introduction),
clearly the $tW$ channel is able to give better bounds on $F_{R2}$.

\begin{table}
\begin{center}
\begin{tabular}{|c|c|c|}\hline
   Systematics  & $F_{L2}$                         & $F_{R2}$  \\ \hline
   $0\%$        &   $-0.039 \leq F_{L2} \leq 0.042$  & $-0.026 \leq F_{R2} \leq 0.017$ \\\hline
   $10\%$        &  $ -0.061 \leq F_{L2} \leq 0.070$  & $-0.040 \leq F_{R2} \leq 0.022 $\\\hline
   $25\%$        &  $ -0.11 \leq F_{L2} \leq 0.15$  & $-0.105 \leq F_{R2} \leq 0.041$ \\
   \hline
   \end{tabular}\label{bounds}
\end{center}\caption{The $68\%$ confidence level bounds on the anomalous couplings assuming  
different values for systematic uncertainties.}
\end{table}

\section{Conclusion}
The $tW$ channel single top quark production at LHC was considered
 as a probe for non-SM couplings in the top quark sector. 
The $tW$ channel provides the
possibility to study the $Wtb$ vertex without receiving
contamination from FCNC. The
transverse momentum distribution of the b-quark from top is almost
insensitive to anomalous couplings while the presence of
anomalous couplings leads to a shift in
transverse momentum of the charged lepton from top (toward the
large transverse momentum region). Using a proposed strategy at
parton level for separation of signal from backgrounds with 20 fb$^{-1}$ of integrated luminosity of data, the
68$\%$ C.L. limits on anomalous couplings (including a combined systematic uncertainty of $25\%$)
are found to be: $-0.11 \leq F_{L2} \leq 0.15~,~-0.105 \leq F_{R2} \leq 0.041$. Comparing
with the limits estimated using the combination of $t-$channel
and $s-$channel of single top quark production in \cite{lev1}, there is more
sensitivity to $F_{R2}$ in this channel and better bounds is
obtained.

{\large \bf Acknowledgments}\\
The author would like to thank CompHEP authors, specially E.
Boos, L. Dudko and A. Sherstnev. Thanks to S. Paktinat for useful
discussions.\\


\begin{thebibliography}{100}
\bibitem{beneke} M. Beneke {\it et al.,} hep-ph/0003033.
\bibitem{tait1} T. M. P. Tait, C.-P. Yuan, Phys. Rev. D63, 014018 (2000).
\bibitem{zhang1} C. J. C. Burgess and H. J. Schnitzer, Nucl. Phys. B228, 454 (1983); W. Buchmuller and
 D. Wyler, Nucl. Phys. B268, 621 (1986).
\bibitem{zhang2} K. Whisnant, J. M. Yang, B. L. Young, X. Zhang, Phys. Rev. D56, 467 (1997).
\bibitem{onfre1} J. A. Aguilar-Saavedra, J. Carvalho, N. Castro, A. Onofre,
 F. Veloso, Eur. Phys. J. C50, 519 (2007).
\bibitem{onfre2} J. A. Aguilar-Saavedra, J. Carvalho, N. Castro, A. Onofre, F. Veloso, arXiv:0705.3041 [hep-ph]
\bibitem{tsuno} S. Tsuno, I. Nakano, R. Tanaka, Y. Sumino,  Phys. Rev. D73, 054011 (2006).
\bibitem{singletopd0} D0 Collaboration, V. M. Abazov, {\it et al.,}, Phys. Rev. Lett. 98, 181802 (2007).
\bibitem{lev1} E. Boos, L. Dudko, T. Ohl, Eur. Phys. J. C11, 473 (1999).
\bibitem{pdg} W.-M. Yao {\it et al.,}, J. Phys. G33, 1 (2006).
\bibitem{cdfwidth} CDF Collaboration, CDF note 8953, August 10, 2007.
\bibitem{kane} G. L. Kane, G. A. Ladinsky, C.-P. Yuan, Phys. Rev. D45, 124 (1992).
\bibitem{mojtaba} M. Mohammadi Najafabadi, J. Phys. G34, 39 (2007).
\bibitem{comphep} E. Boos, V. Bunichev, M. Dubinin, L. Dudko, V. Ilyin, A. Kryukov, V. Edneral,
 V. Savrin, A. Semenov, A. Sherstnev, Nucl. Instrum. Meth. A534, 250 (2004).
\bibitem{cmstdr} CMS Collaboration, CMS PTDR, Vol.II, CERN/LHCC 2006-021, J. Phys. G34, 995 (2007).
\bibitem{atlastdr} ATLAS Collaboration, ATLAS PTDR, Vol.II, CERN/LHCC 1999-15.
\bibitem{cmstdr1} CMS Collaboration, CMS PTDR, Vol.I, CERN/LHCC 2006-001.
\bibitem{tait2} T. M. P. Tait, Phys. Rev. D61, 034001 (1999).
\end{thebibliography}
\end{document}